\documentclass[aps,prd,onecolumn,groupedaddress,nofootinbib,amssymb]{revtex4}
\usepackage{bm}

\usepackage{amsmath}
\usepackage{amssymb}
\usepackage{amsfonts}
\usepackage{wrapfig}
\usepackage{cancel} 
\usepackage[dvipdfmx]{graphicx}
\usepackage{color}
\newcommand{\be}{\begin{equation}}
\newcommand{\ee}{\end{equation}}
\newcommand{\bea}{\begin{eqnarray}}
\newcommand{\eea}{\end{eqnarray}}
\newcommand{\beaa}{\begin{eqnarray*}}
\newcommand{\eeaa}{\end{eqnarray*}}

\allowdisplaybreaks[4]

\begin{document}

\title{Towards construction of ghost-free higher derivative
gravity from bigravity}

\author{Satoshi Akagi$^{1}$\footnote{
	E-mail address: {s\_akagi1108@yahoo.co.jp}   }}

\affiliation{
$^1$ Department of Physics, Nagoya University, Nagoya
464-8602, Japan}

\begin{abstract}
In this paper, the ghost-freeness of the higher derivative theory proposed by Hassan {\it et al.} in [Universe 1 (2015) 2, 92] is investigated. Hassan {\it et al.} believed the ghost-freeness of the higher derivative theory based on the analysis in the linear approximation. However, in order to obtain the complete correspondence, we have to analyze the model without any approximations. In this paper, we analyze two scalar model proposed in [Universe 1 (2015) 2, 92] with arbitrary non-derivative interaction terms. In any order with respect to {perturbative}  parameter, we prove that we can eliminate the ghost for the model with any non-derivative interaction terms.
\end{abstract}


\maketitle
\section{INTRODUCTION}
The question whether the gravity could have the small mass or not has been argued from long time ago. In 1939, M. Fierz and W. Pauli derived the wave equations describing the second order tensor corresponding to massive spin-2 fields, {which} is called the Fierz-Pauli (FP) model \cite{Fierz:1939ix}. Although their works were purely based on field theoretical motivations, some questions began with a negative observation by discovering the vDVZ discontinuity in 1970 \cite{vanDam:1970vg}. The vDVZ discontinuity means that some observables calculated by the Fierz-Pauli theory {do} not coincide with {those} of the massless theory in the massless limit. From the fact, it seems that the possibility of the non-vanishing graviton mass had been excluded. On the other hand, in 1972, A. I. Vainshtein considered the gravitational model where the FP mass terms are added to the Einstein action \cite{Vainshtein:1972sx}. He found that the spherical symmetric solution of the model does not have the discontinuity in the massless limit. Then, it had been obvious that, because the vDVZ discontinuity relies on the linear approximation, the discontinuity could be avoided by considering the non-linear model. This mechanism is called the Vainshtein mechanism. In 1974, however, D. G. Boulware and S. Deser have pointed out that the large class of the massive spin-2 models with non-linear terms, which include the model considered by Vainshtein, has a scalar mode in addition to the massive spin-2 modes \cite{Boulware:1974sr}. This scalar mode has the kinetic term with the negative signature, so called the BD ghost. Then, it {has }become clear that the model is no longer unitary due to the BD ghost. The model satisfying both  the Vainshtein mechanism and the BD ghost-freeness {had} not been constructed for a long time.

The situations changed in 2010. C. de Rham and G. Gabadadze considered the consistency of the non-linear model in the high energy limit, so called {the} decoupling limit. By tuning the parameters of the interaction terms without any derivative, they have obtained the lower-order terms which make the theory ghost-free in the decoupling limit \cite {deRham:2010ik}. After that, they and A. J. Tolley have obtained the full non-linear completion of the non-derivative interaction terms \cite{deRham:2010kj}. Now, this model is called the dRGT model. 
Although they {had} not completed the proof of the absence of the BD ghost in the full non-linear level, S. F. Hassan and R. A. Rosen gave the complete proof by using the Hamiltonian analysis \cite {Hassan:2011hr,Hassan:2011ea}. 
On the other hand, although the dRGT model includes a fixed metric $\eta_{\mu \nu}$ in addition to the dynamical metric $g_{\mu \nu}$ due to the violation of the diffeomorphism, the extension of the flat metric $\eta_{\mu \nu}$ to the general reference metric $f_{\mu \nu}$ {was} investigated. S. F. Hassan {\it et al.} have proved the BD ghost-freeness of the dRGT model with the general reference metric in \cite {Hassan:2011tf}. In addition to the proof, S. F. Hassan and R. A. Rosen have considered a model where the reference metric $f_{\mu \nu}$ becomes dynamical by adding kinetic terms $\sqrt{-f} R(f)$ to the dRGT action, which is called bigravity model. Then, two metrics in this theory have already been interacted with each other. They have proved the BD ghost-freeness of the bigravity model, and they also {showed} that the bigravity includes one massless spin-2 modes and one massive spin-2 modes \cite {Hassan:2011ea,Hassan3}.

On the other hand, the theory which includes the modes identical with the modes in the bigravity has been {also} known in the context of the higher curvature theories. According to \cite{Stelle 1977}, the action where the general second order {terms with respect to the curvature} are added to the Einstein-Hilbert action includes one scalar mode and one massive spin-two mode in addition to the massless spin-2 mode. The scalar mode could be eliminated by tuning the parameters. We call the model where the scalar mode is eliminated as ``the R-squared gravity", in this paper. Although the R-squared gravity has the modes similar to the bigravity, there is {an} essential difference between {both of theories}. Although the bigravity theory does not include any ghost, the signatures between the kinetic terms of the two modes in the R-squared gravity are opposite with each other. Therefore the R-squared gravity violate the unitarity of the S matrix. The only exception is given in the 3 dimensional space-time. E. A. Bergshoeff, O. Hohm and P. K. Townsend proposed the R-squared gravity by tuning the parameters {so} that the massive spin-2 mode has healthy propagation. The obtained model is called the New Massive Gravity (NMG) \cite{Bergshoeff 2009}. Although the signature of the kinetic term for the massless spin-2 mode {is negative}, the massless mode does not propagate in the 3 dimensional space-time. {Then, there is some parameter
 region which makes the theory perturbatively ghost-free. Furthermore, the NMG theory does not include the BD ghost, i.e., the NMG theory  has 2 degrees of freedom in non-linear level. For example, the proof using the St\"{u}kelberg trick was given in \cite{Paulos}. 
In this sense, the NMG theory can be regarded as a higher derivative gravity model conserving the unitarity of the S matrix. However, in the context of the AdS/CFT correspondence,  it is well known that there is no parameter region which keep both the unitarity of the NMG theory, with negative cosmological constant, and the positivity of the central charge of its CFT dual \cite{3d}. The negativity of the central charge means the violation of the unitarity in the theory. Hence the bulk unitarity and the boundary unitarity are incompatible with each other. }

In these backgrounds, the relationship between the bigravity and the R-squared gravity {has} been investigated after the discovery of the bigravity. In particular, M.  F. Paulos and A.  J. Tolley showed the equivalence between the bigravity in some limits of parameters and the R-squared gravity \cite{Paulos}. They also obtained some generalizations of the NMG theory without any BD ghosts. 
Moreover, S. F. Hassan, A. Schmidt-May and M. von Strauss tried to investigate the correspondence between the bigravity and the R-squared gravity without any limits of parameters. They proposed a higher derivative theory describing the same dynamics as bigravity under appropriate conditions. They have also shown that the higher derivative theory coincides with the R-squared gravity in small curvature approximation. In this way, they have concluded that the higher derivative theory is a ghost-free completion of the R-squared gravity. This analysis {has been} extended to a higher order in the context of the correspondence between the Weyl Gravity and the Partially Massless Gravity \cite{Hassan2}.

The reason why they believed the ghost-freeness of the higher derivative theory {was} based on the analysis in the linear approximation with respect to the fields, as given in the Appendix of \cite{Hassan1}. However, in order to obtain the complete correspondence, we have to analyze the model without any approximations. 

In this paper, we analyze the scalar model proposed in \cite{Hassan1} with arbitrary non-derivative interaction terms, and investigate the possibility of the elimination of  {the} ghost.
As a result, we prove it for any non-derivative interaction terms, and any order with respect to the perturbative parameter.

\section{PREVIOUS RESEARCH }
In this section, we briefly review the analysis given in \cite{Hassan1}.
The action of the bigravity model \cite{Hassan3} is given by 
\begin{align}
S[g,f] &= M_g^{D-2} \int d^D x [\sqrt{-g} R(g) + \alpha^{D-2} \sqrt{-f} R(f) 
-2m^2\sqrt{-g} \sum_{n=0}^{D} \beta_n e_n (S) ], \notag \\
e_n(S) &\equiv \frac{1}{n!} 
\delta^{\mu_1 \ \mu_2 \cdots \mu_n}_{ \ \ \nu_1 \ \nu_2 \cdots \nu_n}
S^{\nu_1}_{ \ \ \mu_1} S^{\nu_2}_{ \ \ \mu_2} \cdots S^{\nu_n }_{ \ \ \mu_n} , \ \ 
S^\mu_{ \ \ \nu} \equiv {\sqrt{g^{-1}f}}^\mu_{ \ \ \nu }
, S^\mu_{ \ \ \nu}S^{\nu}_{ \ \ \rho} = g^{\mu \nu}f_{\nu \rho}. \label{bi1}
\end{align}
Here, $D$ is the space-time dimension, $M_g$ is the Planck mass for the metric $g$, $\alpha\equiv M_f/M_g$ {($M_f$ is the Planck mass for the metric $f$.)} is the ratio of the Planck masses, $\beta_n$ are free parameters without dimension, and $m^2$ is the mass parameter, which is introduced in order to make $\beta_n$ dimensionless. {The tensor $\delta^{\mu_1 \ \mu_2 \cdots \mu_n}_{ \ \ \nu_1 \ \nu_2 \cdots \nu_n}$ is defined as follows,
\begin{align}
\delta^{\mu_1 \ \mu_2 \cdots \mu_n}_{ \ \ \nu_1 \ \nu_2 \cdots \nu_n} 
\equiv
\frac{-1}{(D-n)!} \epsilon^{\mu_1 \mu_2 \cdots \mu_n \sigma_{n+1} \cdots \sigma_D } \epsilon_{\nu_1 \nu_2 \cdots \nu_n \sigma_{n+1} \cdots \sigma_D }.
\end{align}
Here the tensor $\epsilon^{\mu_1 \cdots \mu_D}$ is the Levi-Civita anti-symmetric tensor, one of whose components is given by $\epsilon^{012\cdots D-1}$ =1.}
In the action (\ref{bi1}), two metrics $g_{\mu \nu}$ and $f_{\mu \nu}$ interact with each other through the non-derivative interaction terms. Then the equation of motion given by the variation of $g_{\mu \nu}$ does not include any derivative of the metric $f_{\mu \nu}$, 
\begin{align}
\frac{\delta S[g,f]}{\delta g_{\mu \nu}}=0. \label{bi2} 
\end{align}
Therefore we can algebraically solve this equation with respect to $f_{\mu \nu}$. 

The obtained solution depends on the metric $g_{\mu \nu}$ and the curvature $R(g)$.
Although there are generally $D-1$ solutions {in Eq.(\ref{bi2})}, we choose one of them {and denote} as $f_{\mu \nu} [g]$. \footnote[1]{{In \cite{Hassan1}, this difference of the solutions is expressed by a parameter $``a"$, which is a solution of $D-1$ dimensional polynomial equation. In later argument by using scalar fields, we adopt the specific solution. At present, however, we do not restrict the solution. The arguments after Eq.(\ref{bi4})  are correct for any solution.}}
By substituting this solution, {$f_{\mu \nu} = f_{\mu \nu}[g]$}, to the equation of motion obtained by variation with respect to $f_{\mu \nu}$, 
\begin{align}
\left[ \frac{\delta S[g,f] }{\delta f_{\mu \nu}} \right]=0, \label{newe4} 
\end{align}
we obtain,
\begin{align}
\left[ \frac{\delta S[g,f] }{\delta f_{\mu \nu}} \right]_{f=f[g] }=0. \label{bi4}
\end{align}
{This equation of motion (\ref{bi4}) has the following properties: 
The algebraic solution $f_{\mu \nu}[g]$ in (\ref{bi2}) includes the second order derivatives of $g$, and the equation (\ref{newe4}) is second order derivative equation.
Then we find that the equation obtained by substituting the solution $f_{\mu \nu}[g]$ in (\ref{bi2}) into (\ref{newe4}) is the fourth order equation (\ref{bi4}).
Moreover, the solutions of this equation (\ref{bi4}) are also the solutions of the original equations (\ref{bi2}) and (\ref{newe4})},
so that the dynamics described by the equation (\ref{bi4}) are stable despite being the fourth order.

In \cite{Hassan1}, Hassan {\it et al.} proposed the higher derivative model obtained by substituting the algebraic solution $f_{\mu \nu} [g]$ into the original action (\ref{bi1}),
\begin{align}
S'[g]\equiv S[g, f [g]]. \label{bi3}
\end{align}
The dynamics described by this action $S'[g]$ do not completely coincide with the dynamics described by the original action $S[g,f]$. Indeed, by {the} variation of the action (\ref{bi3}), we obtain
\begin{align}
\frac{\delta S'[g]}{\delta g_{\mu \nu} (x)} &
=\left[\frac{\delta S[g,f]}{\delta g_{\mu \nu} (x)}\Biggl|_f \right]_{f=f[g]} +
\int d^D y \frac{\delta f_{\rho \sigma} [g(y)]}{\delta g_{\mu \nu} (x)}
\left[\frac{\delta S[g,f]}{\delta f_{\rho \sigma}(y)} \Biggl|_g \right]_{f=f[g]} \notag \\
&= \int d^D y \frac{\delta f_{\rho \sigma}[g(y)]}{\delta g_{\mu \nu} (x)}
\left[\frac{\delta S[g,f]}{\delta f_{\rho \sigma} (y)} \Biggl|_g \right]_{f=f[g]}.\label{e4}
\end{align}
In the second line, we use the fact that the first term in the first line identically vanishes due to the fact that the algebraic solution $f[g]$ satisfies the equation of motion (\ref{bi2}).
Here, if we define,
\begin{align}
\frac{\delta f_{\rho \sigma}[g(y)]}{\delta g_{\mu \nu} (x)}\equiv \mathcal{O}^{\mu \nu}_{ \ \ \ \rho \sigma}\delta(x-y),
\end{align}
the operator $\mathcal{O}$ becomes the second order derivative operator because the function $f[g]$ contains $R(g)$.
As a result, the equation of motion of {the theory with the action $S'[g]$ in (\ref{bi3})} is given by
\begin{align}
\mathcal{O}^{\mu \nu}_{ \ \ \ \rho \sigma}\left[\frac{\delta S[g,f]}{\delta f_{\rho \sigma} (x)} \Biggl|_g \right]_{f=f[g]}=0. \label{bi5}
\end{align}
{Because the second order differential operator $\mathcal{O}$ acts on the lhs of the original equation (\ref{bi4}) and Eq.(\ref{bi4}) is the fourth order with respect to derivatives, we obtain the 6-th order differential equation (\ref{bi5}).}
By introducing the auxiliary field $\lambda^{\rho \sigma}$, the equation (\ref{bi5}) could be decomposed as follows,
\begin{align}
\mathcal{O}^{\mu \nu}_{ \ \ \ \rho \sigma} \lambda^{\rho \sigma} =0 , \ \ \ \ \ \
\left[\frac{\delta S[g,f]}{\delta f_{\rho \sigma} (x)} \Biggl|_g \right]_{f=f[g]}= \lambda^{\rho \sigma}. \label{bi9}
\end{align}
These equations express the system where the field $\lambda^{\rho \sigma}$ described by the second order differential equation and the field $g_{\mu \nu }$ described by the fourth order differential equation interact with each other.
In order that the solution described by these {equations (\ref{bi9})} is equivalent to the solution described by the original equations {(\ref{bi4})}, it is necessary to be $\lambda^{\rho \sigma}=0$ by choosing the initial conditions and/or the boundary conditions for $\lambda^{\rho \sigma}$.

When we obtain the action (\ref{bi3}), we need to solve the equation (\ref{bi2}) for $f_{\mu \nu}$ explicitly.
It is not, however, so easy to solve the equation (\ref{bi2}) because the equation (\ref{bi2}) is non-linear matrix equation. Then, Hassan {\it et al.} have solved this equation perturbatively by expanding this equation with respect to $1/m^2$. As a result, by substituting the obtained solution {into} the original action (\ref{bi1}), they have shown, 
\begin{align}
S[g, f(g)] = M_g^{D-2} \int d^D x \sqrt{-g} \left[ \Lambda + c_R R(g) -\frac{c_{RR}}{m^2} 
\left( R^{\mu \nu} R_{\mu \nu} -\frac{D}{4(D-1)} R^2 
\right) \right] + \mathcal{O}\left(\frac{1}{m^4} \right).
\end{align}
Here, although the coefficients $\Lambda,c_R,c_{RR}$ are defined by the parameters in the bigravity $\alpha , \beta_n$ in (\ref{bi1}), because {the explicit} forms are a little bit complicated, we do not give these forms now.
By neglecting the higher order terms $\mathcal{O}(\frac{1}{m^4})$, the remaining terms are {those in} the R-squared gravity, which contains the healthy massless spin-2 mode and the ghost-like massive spin-2 mode [for example, see \cite{Paulos,Bergshoeff 2009,Stelle 1977}]. 
Hassan {\it et al.} {have} conjectured that, although the truncated model, which contains the R-squared gravity, includes the ghost, but the complete form of this higher derivative theory could be ghost-free.

The reason why they believed that the higher derivative theory could be ghost-free is based on the analysis in the linear approximation with respect to the fields, as given in the Appendix of \cite{Hassan1}. For avoiding the complication of our argument, we do not use their argument now. Their arguments are given in the Appendix \ref{aa}, which we should read after the argument in section \ref{tc}.

In order to obtain the complete correspondence, we have to analyze {the theory} without any approximations.
Then, we consider the {two scalar} model proposed in \cite{Hassan1}, keeping the interaction terms not equal to zero, 
\begin{align}
S_0[\phi,\psi]= \int d^Dx \left[ \frac{1}{2} \phi \Box \phi + \frac12 \psi \Box \psi 
- \frac{m^2}{2} \left( \phi +\psi \right)^2 - k V(\phi, \psi ) \right], \label{newe1}
\end{align}
and we investigate the possibility {of eliminating the ghost}. We define the function $\psi[\phi]$ as an algebraic solution with respect to $\psi$ of the equation of motion obtained by variation with respect to $\phi$. We consider the higher derivative model obtained by {substituting} the solution $\psi = \psi[\phi]$ to the original action $S_0[\phi,\psi]$ {in (\ref{newe1})},
\begin{align}
S'[\phi] \equiv S_0[\phi,\psi[\phi]]. \label{newe2}
\end{align}
We show that the amplitudes described by $S'[\phi]$, by choosing the appropriate physical space, coincide  with the amplitudes of the original theory.

\section{MODEL OF SCALAR FIELDS}\label{111}
In this section, we give the fundamental properties of the model proposed in this paper.
Because most of the analysis is focused on the linear level, the obtained results are not so different from those obtained by Hassan {\it et al.} \cite{Hassan1} 
but the {spectrum} is obtained by using the formulations different from those in \cite{Hassan1}. 
\subsection{Model of Scaler Fields}
Although the model proposed in this paper has two modes with positive kinetic terms, the corresponding higher derivative model contains an additional mode. 
In this section, we explain this fact by focusing our analysis to the linear terms of fields. 
Let us consider the model of two scalar fields interacting with each other by a mass mixing, 
\begin{align}
S_0 [\phi,\psi]= \int d^D x \left[ \frac12 \phi \Box \phi + \frac12 \psi \Box \psi 
-\frac{m^2}{2} (\phi + \psi)^2 - k V(\phi, \psi) \right]. \label{sc1}
\end{align}
In the analogy with the linearization of the bigravity action (\ref{bi1}) {which includes} the mass mixing 
terms (see \cite{Hassan3}),
 we add the mass mixing term.
We assume that $V(\phi , \psi)$ is the interaction term including the third order or higher powers of fields without derivatives.
This assumption is based on not only the analogy of the non-derivative interaction terms in bigravity, but also the necessity of {expressing $\psi$ as an algebraic function of $\phi$, $\psi=\psi[\phi]$}. 
More generally, although we should add some self interaction terms with some derivatives of $\phi$ and $\psi$ to the action (\ref{sc1}), we do not include them just for simplicity.

Under the field redefinition, 
\begin{align}
\phi = \frac{1}{\sqrt{2}} (\xi + \eta) , \psi= \frac{1}{\sqrt{2}} (\xi- \eta ), \label{mm3}
\end{align}
the linear terms of the action (\ref{sc1}) {are} diagonalized as follows, 
\begin{align}
S_0 [\phi(\xi,\eta)  , \psi (\xi, \eta) ] |_{\text{linear}}
=  \int d^D x \left[ \frac12 \xi (\Box-2m^2) \xi 
+ \frac12 \eta \Box \eta 
 \right].
\label{ssc1}
\end{align}
 From the above expression, we find 
this model includes two scalar fields with mass $0$ and $2m^2$.

By {the} variation of the action (\ref{sc1}) with respect to $\phi$, we obtain
\begin{align}
\frac{\delta S_0}{\delta \phi } = ( \Box - m^2 )  \phi - m^2 \psi
 -k \frac{\partial V (\phi ,\psi )}{ \partial \phi } =0. \label{x1}
\end{align}
Because this equation does not include any derivatives of  $\psi$, we can solve (\ref{x1}) with respect to $\psi$ algebraically.
Because Eq.~(\ref{x1}) is the polynomial with respect to $\psi$, the solution is not unique and the number of the solutions 
depends on the exponent of the highest power terms of $\psi$. 
Nevertheless, the solution corresponding to the vacuum $\phi=0=\psi$ is uniquely determined, 
due to the existence of the mass mixing term.
Then, {just for simplicity,} we adopt $\psi[\phi]$ which is the algebraic solution satisfying $\psi[\phi=0] =0$.\footnote[2]{{In this paper, we analyze the scattering amplitudes in order to argue the ghost-freeness of the higher derivative theory (\ref{newe2}). When we calculate the perturbative scattering amplitude, we have to choose the vacuum. As will be discussed later, the amplitudes of the higher derivative theory with the assumption $\psi[\phi=0] =0$ correspond to those of the original theory calculated around the vacuum $\phi=0=\psi$. On the other hand, another algebraic solution corresponds to the perturbative theories around another vacuum. We can extend our discussion to including another vacuum by replacing the mass terms $m^2(\phi+\psi)^2$ in the action (\ref{sc1}) with more general terms $m_\phi^2 \phi^2 +2 m_{\phi \psi}^2 \phi \psi +  m_\psi^2 \psi^2 $ (and regarding the replaced action as the action which have already been expanded around the interested vacuum). In the argument under the replaced action, the assumption $\psi[\phi=0] =0$ is no longer the specific case. But, just for simplicity, we do not extend the discussion in this paper.}} {Through all of the later arguments in this paper, we adopt this class of solution.}
In this assumption, the linear part of $\psi[\phi]$ is expressed as follows, 
\begin{align}
m^2 \psi[\phi] = (\Box -m^2)\phi + \mathcal{O}(\phi^2) .\label{sc3}
\end{align}

The higher derivative theory derived by Hassan {\it et al.} \cite{Hassan1} corresponds to the new action $S_0[\phi,\psi[\phi]]$
 obtained by substituting the algebraic solution (\ref{sc3}) {into} the original action (\ref{sc1}).
Then the linear terms in the action are expressed as follows, 
\begin{align}
S_0 [\phi, \psi[\phi]] =  \int d^D x \left[ \frac{1}{2m^4} \phi \Box (\Box - 2m^2) (\Box -m^2 ) \phi  
 + \mathcal{O}(\phi^3)\right]. \label{sc5}
\end{align}
From the action (\ref{sc5}), we find that there are the mass spectrum $0$, $2m^2$, and an additional spectrum $m^2$.
In the next part, we investigate whether each of the modes {is} ghost or not.

\subsection{{Spectrum}}\label{ms}
We have found the higher derivative model, whose linear parts are given in (\ref{sc5}), 
contain an additional field with mass squared $m^2$.
Now, in order to check whether the additional spectrum and original modes {are} ghost or not,
we decompose {the action (\ref{sc5})} by introducing the 
Lagrange multiplier field $\lambda$,  
\begin{align}
S_0[\phi, \psi[\phi]] \longrightarrow 
S [\phi, \psi , \lambda] \equiv S_0[\phi, \psi] + m^2 \int d^D x \lambda \left( \psi -\psi[\phi] \right) . \label{gg5}
\end{align}
Here, in order to simplify the later arguments, we put the coefficient $m^2$ in front of $\lambda$.
Indeed, this coefficient does not affect the dynamics no {matter} how we choose it.
By using the equation (\ref{sc3}), the linear part of $S [\phi, \psi , \lambda]$ can be rewritten as follows, 
\begin{align}
S [\phi , \psi , \lambda] |_{\text{linear}} &= \int d^D x
\left[ \frac12 \phi \Box \phi 
+ \frac12 \psi \Box \psi - \frac{m^2}{2} (\phi + \psi)^2 
+  \lambda \left(m^2 \psi -  ( \Box - m^2 ) \phi \right) \right] \notag \\
&= S_0[\phi,\psi] |_{\text{linear}} - \int d^D x \lambda (x)  
\frac{\delta S_0[\phi,\psi] |_{\text{linear}}}{\delta \phi (x)}.
\end{align}
From the last line of {the} above equations, it is obvious that the Lagrange multiplier terms could be eliminated by the field redefinition $\phi \longrightarrow \phi + \lambda$,
\begin{align}
S [\phi + \lambda , \psi , \lambda] |_{\text{linear}} = \int d^D x \left[ \frac12 \phi \Box \phi 
+ \frac12 \psi \Box \psi - \frac{m^2}{2} (\phi + \psi)^2 
- \frac12 \lambda ( \Box -m^2 ) \lambda \right].\label{gg1} 
\end{align}
Then we find that $\lambda$ has the kinetic term with a negative signature;
therefore,  $\lambda$ is a ghost with mass $m^2$.  
On the other hand, we also find that the terms of $\phi$ and $\psi$ correspond to the linear terms of the original theory (\ref{sc1}); therefore, $\phi$ and $\psi$ are healthy fields.
Indeed, under the field redefinitions,
\begin{align}
\phi = \frac{1}{\sqrt{2}} (\xi + \eta) , \psi= \frac{1}{\sqrt{2}} (\xi- \eta ),
\label{xieta}
\end{align}
we obtain following diagonal expression, 
\begin{align}
&\bar{S}[\xi,\eta,\lambda] |_{\text{linear}}
=  \int d^D x \left[ \frac12 \xi (\Box-2m^2) \xi 
+ \frac12 \eta \Box \eta 
- \frac12 \lambda ( \Box -m^2 ) \lambda \right], \notag \\
&\bar{S}[\xi,\eta,\lambda]\equiv S [\phi(\xi,\eta) + \lambda , \psi (\xi, \eta) , \lambda] {.}  \label{sc4}
\end{align}
This action (\ref{sc4}) has the healthy modes $\xi$ and $\eta$, and the extra ghost mode $\lambda$.
In other words, although the fields included in the original action $S_0[\phi,\psi]$ (\ref{sc1}) are healthy, 
the new ghost field has appeared by the procedure of substitution. 
In the following sections, we would like to call $(\xi,\eta)$ in the action (\ref{sc4}) or $(\phi, \psi)$ in the action (\ref{gg1}) ``physical fields.''

\section{CONJECTURE AND SAMPLE CALCULATION}\label{ex}
The purpose of this section is to conjecture the correspondence between the original theory $S_0[\phi, \psi]$ (\ref{sc1}) 
and the corresponding higher derivative theory $S_0[\phi,\psi[\phi]]$ (\ref{sc5}).
For this purpose, we consider specific interaction terms and calculate the tree-level amplitudes of the higher derivative theory described by the action $S_0[\phi,\psi[\phi]]$ (\ref{sc5}).

\subsection{The Conjecture}\label{tc}

As a result, ``physical amplitudes'' of the higher derivative theory coincide with the scattering amplitudes of the original theory. Here, ``physical amplitudes'' mean the scattering amplitudes {where} all external lines are taken to physical fields 
$\xi$ and $\eta$ in (\ref{xieta}). 
In other words, the conjecture is expressed as the realization of the correspondence, 
\begin{align}
&\langle \xi(\vec{k_1} )  \cdots \xi(\vec k_n ) \eta(\vec k_{n+1}) \cdots \eta(\vec k_{n+m}) ; out 
| \xi(\vec{p_1} )  \cdots \xi(\vec p_N) \eta(\vec p_{N+1}) \cdots \eta(\vec p_{N+M}) ; in \rangle_\mathrm{Original} 
\notag \\ 
&= \langle \xi(\vec k_1  )  \cdots \xi(\vec k_n ) \eta(\vec k_{n+1}) \cdots \eta(\vec k_{n+m}) ; out 
| \xi(\vec p_1 )  \cdots \xi(\vec p_N) \eta(\vec p_{N+1}) \cdots \eta(\vec p_{N+M}) ; in \rangle_\mathrm{HD}, \label{ex1}
\end{align}
in the tree-level.
{As will be discussed later}, because the Green functions of both theories are not identical with each other, 
we express the correspondence by using {the} S matrix elements.   
This means that the correspondence is only valid under the on-shell conditions.
The reason why we find this conjecture 
and the proof for the specific case are given in the Appendix \ref{aa} .

\subsection{A Sample Calculation}\label{asc}
\begin{figure}
\centering
\includegraphics[width=13cm]{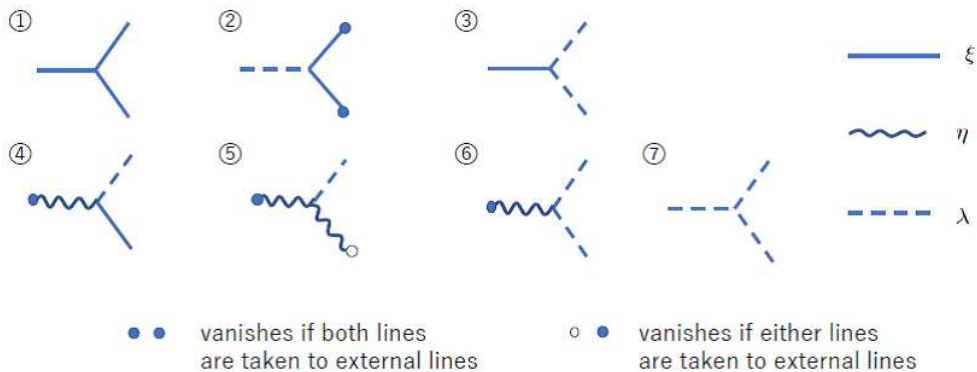}
\caption{Interaction terms in the third order}
\label{3rd}
\end{figure}

In this part, in order to confirm the validity of the conjecture (\ref{ex1}), we {investigate} the structure of the Feynman diagrams of {the} higher derivative theory for given interaction terms.
As a result, we find the interesting structure of the diagrams.  
Now, we consider the third order interaction term with respect to the massive field $\xi$,
\begin{align}
S[\xi,\eta] &= \int d^D x \left[ \frac12 \xi (\Box -2m^2) \xi + \frac12 \eta \Box \eta 
- \mu \frac{\sqrt{2}}{3} \xi^3  \right] \notag \\
&= \int d^D x \left[ \frac12 \phi \Box \phi +\frac12 \psi \Box \psi 
- \frac{m^2}{2} (\phi+ \psi)^2 
 -\frac{\mu }{3!} 
(\phi+\psi)^3  \right].
\label{3action}
\end{align}
We start with deriving the corresponding higher derivative theory.
The equation of motion obtained by the variation of the action (\ref{3action}) with respect to $\phi$,
\begin{align}
\frac{\delta S}{\delta \phi} = (\Box -m^2) \phi - m^2 \psi {-} \frac{\mu}{2} (\phi+ \psi)^2 =0
\label{eom1}
\end{align}
could be solved for $\psi$ as follows, 
\begin{align}
\psi[\phi]= \phi - \frac{m^2}{\mu} \pm \sqrt{\frac{m^4}{\mu^2} + \frac{2}{\mu} \Box \phi}.
\label{sol1}
\end{align}
We find that there are two solutions because the equation of motion (\ref{eom1}) is quadratic with respect to $\psi$. 
Now we restrict our arguments to the unique solution with the vacuum $\phi=0=\psi$.
The signature satisfying this condition is ``$+$'' in (\ref{sol1}).
Under this selection, Eq.~(\ref{sol1}) can be expanded with respect to $\mu$ as follows,  
\begin{align}
m^2 \psi[\phi] =  (\Box -m^2) \phi   - \frac{\mu}{2m^4} (\Box \phi)^2 
+ \frac{\mu^2}{2m^8} (\Box \phi)^3 + \mathcal{O} (\mu^3). \label{jk1}
\end{align}
By replacing $\psi$ in the original action $S_0[\phi ,\psi]$ with $\psi[\phi]$, we obtain the higher derivative theory.
However, for the simplicity of the analysis, we do not consider the higher derivative form.
Instead of this, we analyze the action (\ref{sc4}) {expressed} by $\xi$, $\eta$, and $\lambda$.
Now we expand the action with respect to $\mu$, 
\begin{align}
S [\phi(\xi,\eta) + \lambda , \psi (\xi, \eta) , \lambda] 
\equiv \sum_{n=0}^\infty \bar{S}^{(n)} [\xi,\eta,\lambda] .
\end{align}
Here $\bar{S}^{(n)} [\xi,\eta,\lambda]$ are the $n$-th order terms with respect to $\mu$.
By using this notation, the lower-order terms of $\bar{S}^{(n)} [\xi,\eta,\lambda]$ corresponding to 
the algebraic solution (\ref{jk1}) are given by
\begin{align}
\bar{S}^{(0)}[\xi,\eta,\lambda] &=   \int d^D x \left[ \frac12 \xi (\Box-2m^2) \xi 
+ \frac12 \eta \Box \eta 
- \frac12 \lambda ( \Box -m^2 ) \lambda \right] , \notag \\
\bar{S}^{(1)}[\xi,\eta,\lambda] &=   \int d^D x \left[  -\frac{\sqrt{2}}{3} \mu \xi^3 
+ \frac{\mu}{4m^2} \lambda \left\{ (\Box \xi)^2 -4m^4 \xi^2 \right\}  
 + \frac{\mu}{\sqrt{2}} \lambda \left\{ \Box \xi \Box \lambda 
-m^4 \xi \lambda  \right\} \right. \notag \\
& \ \ \ \ \ \ \ \ \ \ \ \  \ \ + \frac{\mu}{4m^4} \lambda 
\left\{ 2\Box \xi + \Box \eta +2 \sqrt{2} \Box \lambda \right\} \Box \eta 
-\frac{\mu}{3!} \lambda^3 + \frac{ \mu}{2m^4} \lambda(\Box \lambda)^2 
 \Biggl] , \notag \\
\bar{S}^{(2)}[\xi,\eta,\lambda] &=   \int d^D x 
 \left( -\frac{\mu^2}{2m^8} \right) \lambda \left[ \frac{1}{2\sqrt{2}} \left( \Box \xi +\Box \eta
 \right)^3 + \frac{3}{2} \left( \Box \xi + \Box \eta \right)^2 \Box \lambda \right.  \notag \\ 
& \ \ \ \ \ \  \ \  \ \ \ \ \ \ \ \ \ \ \  \ \ \ \ \ \ \ \ \ \left.
+\frac{3}{\sqrt{2}} (\Box \xi + \Box \eta ) (\Box \lambda)^2 + (\Box \lambda)^3 \right].\label{jk2}
\end{align}
The lowest terms in {$\bar{S}^{(0)}$} coincide with those in (\ref{sc4}).
The Feynman diagrams of the third order terms are summarized in FIG. \ref{3rd}.
Now let us {investigate} the sufficient condition for the realization of the conjecture given in (\ref{ex1}). 
We should note again that the following arguments are only in the case of tree-level.

In the first order terms { with respect to $\mu$, $\bar{S}^{(1)}$}, the first term, which is represented as \textcircled{\scriptsize 1} in FIG. \ref{3rd}, coincides with the interaction term of the original action (\ref{3action}).
For the realization of the conjecture (\ref{ex1}), it is enough to show that the amplitudes including the vertexes except the vertex \textcircled{\scriptsize 1} do not contribute to the physical amplitudes. 
Moreover, because the terms $\bar{S}^{(n)}, \ n\geq {2}$, {include} the Lagrange multiplier field $\lambda$, we find that all terms except \textcircled{\scriptsize 1} always include the ghost $\lambda$. Then, if {any} contribution from some vertexes except \textcircled{\scriptsize 1} exists, there must be the ghost in the internal line.
Therefore, for the realization of (\ref{ex1}), it is enough to show that the diagrams including the ghost in internal line do not contribute to the physical amplitudes.

\begin{figure}
\centering
\includegraphics[width=12cm]{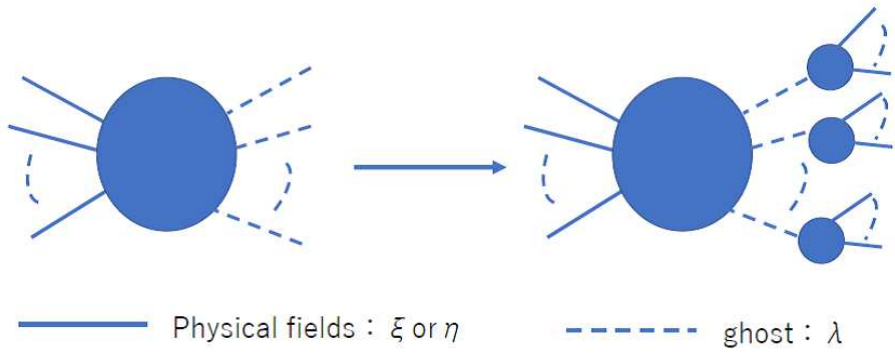}
\caption{The operation decreasing the number of ghost}
\label{gh}
\end{figure}

In order to confirm the sufficient condition above, it is enough to show there are no vertexes which could {decrease} the number of ghosts.
The reason and the more exact meaning of this statement may be obvious by the following consideration:
Let us consider any vertex which contains the some physical fields and some ghost fields, in the left hand side of 
FIG. \ref{gh}.
Here, the solid lines represent the physical fields, and the broken lines represent the ghost.
If we try to construct the physical amplitudes from this vertex, we must decrease the  number of ghosts by acting some vertexes, as in the right hand side of FIG. \ref{gh}.  
The vertex which could decrease the number of the ghosts are only the first-order terms with respect to ghost. Then, if the action {does} not contain the first order terms with respect to the ghost, we cannot construct the diagrams contributing to the physical amplitude from the vertex in the left hand side of FIG. \ref{gh}.  

\begin{figure}
\centering
\includegraphics[width=10cm]{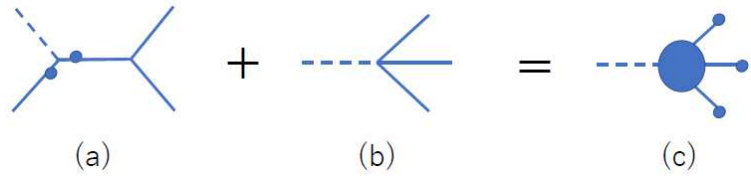}
\caption{The sums of the non-vanishing diagrams}
\label{diag}
\end{figure}

\begin{figure}
\centering
\includegraphics[width=12cm]{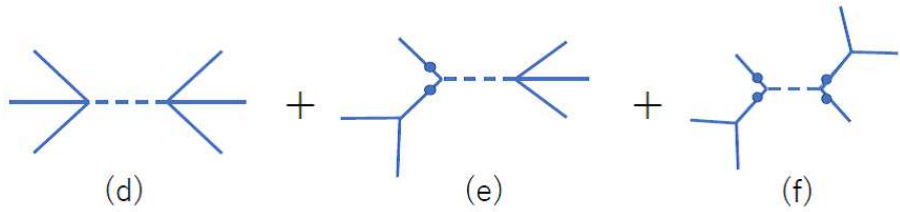}
\caption{The contribution of the non-vanishing diagrams to the physical amplitudes}
\label{amp}
\end{figure}

Although the action (\ref{jk2}) seems to contain the first order terms with respect to the ghost, [that is, the vertexes \textcircled{\scriptsize 2}, \textcircled{\scriptsize 4}, and \textcircled{\scriptsize 5} in FIG. \ref{3rd}], these terms effectively do not contribute to the physical amplitudes.  
The vertex \textcircled{\scriptsize 2} obtained from following terms;
\begin{align}
\frac{\mu}{4m^2} \lambda \left\{ (\Box \xi)^2 -4m^4 \xi^2 \right\},  \label{ex2}
\end{align}
vanishes if both of the physical fields $\xi$ are taken to external line.
Indeed, the on-shell condition {is} given by $\Box \longrightarrow 2m^2$, then the terms  (\ref{ex2}) obviously vanish. 
Similarly, we find that the contributions from \textcircled{\scriptsize 4} and \textcircled{\scriptsize 5} vanish again 
under the on-shell condition of $\eta$.   
In order to emphasize this fact, we put some points on the diagrams represented in FIG. \ref{3rd}. 
The lines which connect the two points with same color vanish by taking all the lines with the points to the external lines 
simultaneously.
From this property, it is impossible to construct the non-vanishing physical amplitude by using the terms (\ref{ex2}) 
at least lower than 6 points diagrams.
The non-triviality {appears} in 6 points diagrams.
Because we can construct the non-vanishing diagrams like as (a) of FIG. \ref{diag} with on-shell condition for the physical fields, without the on-shell condition of $\lambda$, the 6 points diagram like as (f) of FIG. \ref{amp} could survive under the on-shell condition.
So that, we afraid if this diagram could contribute to the physical amplitudes.
In this order, however, we cannot {ignore} the contributions from the higher order terms,
\begin{align}
 \left( -\frac{\mu^2}{2m^8} \right) \lambda \frac{1}{2\sqrt{2}} \left( \Box \xi  \right)^3,
\end{align}
{in $\bar{S}^{(2)}$.} Surprisingly, the summation of the diagrams (a) and (b) with the on-shell condition of the physical fields, without the on-shell condition of $\lambda$, {becomes equal to zero}. 
So we could regard the summation of these diagrams as the diagram (c) of FIG. \ref{diag}.
Therefore, under the on-shell condition, the summation of the non-vanishing amplitudes represented in FIG. \ref{amp} {becomes equal to zero}.
In this way, we predict that the non-vanishing diagrams constructed from the lower order terms could be eliminated by the diagrams constructed from the higher order terms.
In the following sections, we would like to prove this conjecture for general non-derivative interaction terms in any order.


\clearpage

\section{GENERAL PROOF}\label{333}
In {the} previous sections, we specified the modes of the theory described by the action $S_0[\phi,\psi[\phi]]$, and obtained the conjecture for the correspondence between the higher derivative theory $S_0[\phi,\psi[\phi]]$ and the original theory $S_0[\phi,\psi]$.
In this section, we would like to argue general non-derivative interaction terms $V(\phi,\psi)$, and prove the conjecture in any order with respect {to} the perturbative parameter $k$. 

\subsection{{General Algebraic Solution}}
First, in this section, we derive the lower order terms with respect to $k$ in $\psi[\phi]$.
The full-order solution will be derived in a later section.

Let us consider the action (\ref{sc1}) with general non-derivative interaction terms,
\begin{align}
S_0= \int d^Dx \left[ \frac{1}{2} \phi \Box \phi + \frac12 \psi \Box \psi 
- \frac{m^2}{2} \left( \phi +\psi \right)^2 - k V(\phi, \psi ) \right].
\end{align}
Here,  $V(\phi , \psi)$ consists of the general non-derivative interaction terms including third
or higher order terms of fields.
The equation of motion derived by {the variation with} respect to $\phi$,
\begin{align}
&\frac{\delta S_0}{ \delta \phi} = \left( \Box -m^2\right) \phi - m^2 \psi - k V^{1,0}(\phi , \psi)=0,
\notag \\
&V^{n,m}(\phi, \psi) \equiv \frac{\partial^{n+m} V(\phi, \psi)}{\partial^n \phi \partial^m \psi}, \label{kk1}
\end{align}
could be solved {with} respect to $\psi$ around the vacuum $\phi=0=\psi$ for any $V(\phi , \psi)$.
We assume a perturbative solution expanded with respect to $k$ in the following form, 
\begin{align}
m^2 \psi[\phi] =  m^2 \psi_0 [\phi]+ F[\phi] ,  \ \ \ m^2 \psi_0 [\phi] \equiv \left( \Box -m^2\right) \phi 
, \ \ \ 
 F[\phi] \equiv 
\sum_{n=1}^{\infty} k^n F^{(n)}[\phi], \label{kk2}
\end{align}
and determine the $F[\phi]$.
By substituting (\ref{kk2}) {into} (\ref{kk1}), and expanding the obtained expression in powers of $k$, we find,
\begin{align}
- \frac{1}{k}F[\phi] &=V^{1,0}\left( \phi , \psi_0[\phi] + \frac{1}{m^2}F[\phi] \right) \notag \\
&= \sum_{n=0}^{\infty} \frac{V^{1,n} (\phi , \psi_0[\phi])}{n!} \left( \frac{F[\phi]}{m^2} \right)^n \notag \\
&= \sum_{n=0}^{\infty}  \frac{1}{m^{2n}} \frac{V^{1,n}(\phi , \psi_0[\phi])}{n!}
\sum_{k_1 =1}^{\infty} \cdots \sum_{k_n =1}^{\infty} F^{(k_1)} [\phi] \cdots F^{(k_n)}[\phi]
k^{k_1 +\cdots + k_n} \notag \\
&= \sum_{n=0}^{\infty}  \frac{1}{m^{2n}} \frac{V^{1,n}(\phi , \psi_0[\phi])}{n!}
\sum_{N=1}^{\infty}k^{N}
\sum_{1\leq k_1, \cdots , k_n \leq N-n+1 } F^{(k_1)} [\phi] \cdots F^{(k_n)}[\phi]
 \delta_{k_1 + \cdots k_n , N} \notag \\
&=V^{1,0}(\phi , \psi_0) +
\sum_{N=1}^{\infty}k^{N} \sum_{n=1}^{N}  \frac{1}{m^{2n}} \frac{V^{1,n}(\phi , \psi_0[\phi])}{n!}
 \sum_{s_1 =1}^{N-n+1} F^{(s_1)} [\phi] \sum_{s_2 =1}^{N-n+1-s_1} F^{(s_2)} [\phi] \cdots \notag \\ & \ \ \ \ \ \ \ \sum_{s_{n-1} =1}^{N-n-s_1 -s_2 -\cdots -s_{n-2}}  F^{(s_{n-1})}[\phi] F^{(N-n+1-s_1 -s_2 -\cdots - s_{n-1})} [\phi] \notag \\
&= V^{1,0}(\phi , \psi_0 [\phi]) +\sum_{N=1}^{\infty} k^{N} \sum_{n=1}^{N}  \frac{1}{m^{2n}} \frac{V^{1,n}(\phi , \psi_0 [\phi] ) }{n!} \notag \\
& \ \ \ \ \ \ \times \left[ \Pi_{i=1}^{n-1} \sum_{s_i =1}^{N-n+1 - \Sigma_{k =1}^{ i-1}s_k }F^{(s_i)} [\phi]  \right]  
F^{(N-n+1 - \Sigma_{k =1}^{n-1}s_k )} [\phi].
\label{expa1}
\end{align} 
By using the last expression for $F[\phi]$ in (\ref{kk2}) and by 
comparing both sides in (\ref{expa1}) order by order in $k$, we obtain following recursion relations,  
\begin{align}
&F^{(1)} = - V^{1,0}(\phi, \psi_0[\phi]), \notag \\
&F^{(N+1)} = -\sum_{n=1}^{N}\frac{1}{m^{2n}}  \frac{V^{1,n}(\phi, \psi_0[\phi])}{n!}
\left[ \Pi_{i=1}^{n-1} \sum_{s_i =1}^{N-n+1 - \Sigma_{k =1}^{ i-1}s_k }F^{(s_i)} [\phi] \right]  
F^{(N-n+1 - \Sigma_{k =1}^{n-1}s_k )}  [\phi].
\end{align}
Solving the recursion relations for lower order terms, we find 
\begin{align}
&F^{(1)} [\phi] =  - V^{1,0}_0, \notag \\
&F^{(2)} [\phi] =  \frac{1}{m^2}  V^{1,0}_0 V^{1,1}_0, \notag \\
&F^{(3)} [\phi] = -\frac{1}{m^4} V^{1,0}_0 \left[ \left(V^{1,1}_0   \right)^2  
  +\frac{1}{2} V^{1,2}_0  V^{1,0}_0     \right], \notag \\
&F^{(4)} [\phi] = \frac{1}{m^6}  V^{1,0}_0 \left[  \left( V^{1,1}_0 \right)^3 + \frac32 V^{1,0}_0 V^{1,1}_0
V^{1,2}_0 + \frac16 \left( V^{1,0}_0 \right)^2 V^{1,3}_0 \label{ff1}
\right].
\end{align}
Here, we express $V^{n,m}(\phi, \psi_0 [\phi])$ as $V^{n,m}(\phi, \psi_0 [\phi]) \equiv V_0^{n,m}$ for simplicity.

\subsection{Validity of Conjecture}
In this section, we confirm the conjecture (\ref{ex1}) for any non-derivative interaction terms, $V(\phi, \psi )$, in 
lower order perturbations.
As we have seen in Section \ref{ex}, a sufficient condition realizing the conjecture (\ref{ex1}) is that {the action of the higher derivative theory does not include the first-order terms with respect to $\lambda$.}
Then we consider eliminate the first-order terms with respect to $\lambda$ by some field redefinitions.

According to the Kamefuchi-O'Raifeartaigh-Salam's theorem \cite{Kamefuchi}, the S matrix elements are invariant, under the fields redefinitions expressed as follows, 
\begin{align}
\phi' =  c\phi + u[\phi, \psi, \lambda], \label{gg3}
\end{align}
where $c$ must be a non-vanishing constant, and $u[\phi,\psi,\lambda]$ must be second or higher order terms with respect to
fields or some derivatives of these fields.
Then,  it is enough to confirm that all field redefinitions [except for later (\ref{gg2})] satisfy the expression (\ref{gg3}), in order to realize the conjecture (\ref{ex1}).

Let us consider the action obtained by substituting the algebraic solution (\ref{kk2}) {into} the action decomposed by {the} Lagrange multiplier field $\lambda$ (\ref{gg5}), 
\begin{align}
S_1[\phi_1,\psi , \lambda ] &\equiv   \int d^Dx \left[  \frac{1}{2} \phi_1 \Box \phi_1 + \frac12 \psi \Box \psi 
- \frac{m^2}{2} \left( \phi_1 +\psi \right)^2 - k V(\phi_1 , \psi ) +  \lambda(m^2\psi- 
m^2 \psi_0[\phi_1] - F[\phi_1]) \right] \notag \\
&= S_0 [\phi_1,\psi] +  \int d^Dx \left[  \lambda(m^2\psi- 
m^2 \psi_0[\phi_1] - F[\phi_1]) \right].\label{jj1}
\end{align}
Here, in order to regard the successive redefinitions of the action and $\phi$ in the following as some arithmetic progressions, 
we put the number ``1'' as a suffix on them.
As we have seen in Section \ref{ms}, the first-order terms with respect to $\lambda$, in the zero-order of $k$, could be eliminated by the field redefinition,  
\begin{align}
\phi_1 = \phi_2+\lambda . \label{gg2}
\end{align}
Under the redefinition (\ref{gg2}),  the action (\ref{jj1}) is transformed as follows, 
\begin{align}
S_2[\phi_2 ,\psi ,\lambda ] \equiv S_1[\phi_1 , \psi ,\lambda ] &=  S_0 [\phi_2,\psi] 
+  \int d^Dx \lambda \left[ - k V^{1,0} (\phi_2 , \psi) + F[\phi_2]  \right] 
+ \mathcal{O}(\lambda^2) \notag \\
&= S_0 [\phi_2,\psi] 
+  \int d^Dx \lambda \left[ - k \left( V^{1,0} (\phi_2 , \psi) + F^{(1)}[\phi_2] \right) - \sum_{n=2}^{\infty} k^n F^{(n)}[\phi_2]  \right] 
+ \mathcal{O}(\lambda^2) .\label{jj2}
\end{align}
Because we are not interested in second or higher order terms with respect to $\lambda$,
we could {ignore} these terms.

Now, the terms independent of $k$ {have vanished} and new terms proportional to $k$ {have appeared}.  
The new terms are expressed by the second term of the first line in (\ref{jj2}), $- k\lambda V^{1,0} (\phi_2 , \psi)$, 
that is, the contribution from $S_0$.
Then, in the second line of  (\ref{jj2}), we pick up the first-order terms with respect to $k$.
We could show that these terms vanish under the on-shell condition.
By the result in (\ref{ff1}), the first-order terms with respect to $k$
are expressed as follows, 
\begin{align}
-k \lambda \left( V^{1,0} (\phi_2 , \psi ) - V^{1,0} (\phi_2 , \psi_0[\phi_2])  \right). \label{kk3}
\end{align}
The only difference between the two terms is that the arguments are either $\psi$ or $\psi_0[\phi_2]$.
Now, because the linear terms have already been diagonalized, the condition, $ m^2\psi= m^2\psi_0[\phi_2] \equiv (\Box - m^2) \phi_2 $, is the linear parts of the solution of the EoM obtained by the variation of the action respect to $\phi_2$, i.e., the on-shell condition.
Hence, the contributions from the vertexes (\ref{kk3}) to the scattering amplitudes vanish when all the physical fields $\phi_2 ,\psi$ are taken to the external lines. 
These terms just correspond to the terms which vanish under the on-shell condition in the 
$\xi^3$-model,
\begin{align}
\frac{\mu}{4m^2} \lambda \left\{ (\Box \xi)^2 -4m^4 \xi^2 \right\} .
\end{align}
Now, because we have assumed that $V^{1,0}(\phi,\psi)$ includes third or higher order terms of fields, we have verified that the diagrams, less than 6 points, including some internal lines of ghost, {do} not contribute to the physical amplitudes for general non-derivative interaction terms.

Moreover, the terms (\ref{kk3}) could be eliminated by additional field redefinition.
The fact that the terms (\ref{kk3}) become equal to zero under the on-shell condition $\psi = \psi_0 [\phi_2]$, means that {the terms (\ref{kk3})} could be factored by $(\psi -\psi_0[\phi_2])$.
Indeed, by expanding $V^{1,0}(\phi_2 ,\psi)$ around $\psi=\psi_0[\phi_2]$, 
because the leading terms are canceled each other out,  the terms (\ref{kk3}) could obviously be factored by $(\psi -\psi_0[\phi_2])$.
Because the terms $(\psi -\psi_0[\phi_2])$ are the linear part of EoM, we could eliminate {these} terms by some field redefinition.
Indeed, under the field redefinition,
\begin{align}
\phi_2= \phi_3 + \lambda u [\phi_3, \psi ], \label{kk5}
\end{align}
the contributions from $S_0$ are given by
\begin{align}
 S_0 [\phi_2= \phi_3 + \lambda u [\phi_3 , \psi] ,  \psi] - S_0[\phi_3 , \psi ]
 = - \int d^D x  
[m^2 (\psi - \psi_0 [\phi_3]) + k  V^{1,0 }(\phi_3 , \psi)] \lambda u(\phi_3, \psi) 
+ \mathcal{O} (\lambda^2). \label{kk8}
\end{align}
The contributions from the other terms could be included in $\mathcal{O}(\lambda^2)$.
Then (\ref{kk3}) could be eliminated, if we choose the term $u[\phi_3, \psi]$ as follows, 
\begin{align}
&{u[\phi_3 , \psi]} = -\frac{k}{m^2} \frac{\Delta V^{1,0}(\phi_3 , \psi)}{\Delta \psi}, \notag \\
&\Delta V^{1,0}(\phi_3 , \psi) \equiv V^{1,0} (\phi_3 , \psi) - V^{1,0} (\phi_3 , \psi_0[\phi_3] ),
 \ \ \ \Delta \psi \equiv \psi -\psi_0 [\phi_3].  \label{kk4}
\end{align}
We should note that, although the representation (\ref{kk4}) seems to be defined as a division of
$\Delta \psi$, because $\Delta V$ proportional to $\Delta \psi$, it is in fact some polynomial of the fields.
So that, the field redefinition (\ref{kk5}) satisfies the expression (\ref{gg3}), and the S matrix elements are invariant under this field redefinition.

For the convenience of later arguments, we now define the operator $\Delta$ for any function $f(\psi)$ of $\psi$ as follows, 
\begin{align}
\Delta f(\psi) \equiv f(\psi) - \lim_{\psi \longrightarrow \psi_0} f(\psi).
\label{Delta}
\end{align}
We should note that $\Delta V^{1,0} (\phi , \psi)$ and $\Delta \psi$ in (\ref{kk4}) surely satisfy this definition.
Here, for the field redefinitions later, the argument $\phi$ of the limiting value $\psi_0[\phi]$ is defined as the adopted variable $\phi_n$ {for each frame}.
In other {words}, more correctly, we should define the operator $\Delta_n$ as follows,
\begin{align}
\Delta_n f(\psi) \equiv f(\psi) - \lim_{\psi \longrightarrow \psi_0[\phi_n]} f(\psi),
\label{Delta2}
\end{align}
but we ignore the index $n$ just for simplicity.
Indeed, the difference of each index $n$ could be included into the ignored term $\mathcal{O}(\lambda^2)$ in all of the following relevant equations.

After the field redefinition (\ref{kk5}), the action could be expressed as follows, 
\begin{align}
S_3[\phi_3 , \psi ,\lambda ] \equiv& S_2[\phi_2 = \phi_3 + \lambda u[\phi_3 , \psi]
, \psi ,  \lambda ]\notag \\
=& S_0[\phi_3 , \psi] 
+\int d^Dx \left[ \frac{\lambda k^2}{m^2}  \left[
V^{1, 0}(\phi_3 , \psi) \frac{\Delta V^{1,0} (\phi_3 , \psi) }{\Delta \psi } 
- m^2 F^{(2)}[\phi_3] \right]  \right. \notag \\
& \ \ \ \ \  \ \ \ \ \ \ \ \ \ \ \ \ \ \ \ \ \ \ \ \ \ 
 \left. -\lambda \sum_{n=3}^{\infty} k^n F^{(n)} [\phi_3] 
+ \mathcal{O} (\lambda^2 ) \right].
\end{align}
From the result of the previous part, $m^2 F^{(2)} [\phi_3] = V^{1,0}(\phi_3, \psi_0[\phi_3])
V^{1,1} (\phi_3, \psi_0 [\phi_3]) $, we find the terms proportional to second powers of $k$ are given by
\begin{align}
\frac{\lambda k^2}{m^2} \left[
V^{1, 0}(\phi_3 , \psi) \frac{\Delta V^{1,0} (\phi_3 , \psi) }{\Delta \psi } 
-   V^{1,0}(\phi_3, \psi_0[\phi_3]) 
V^{1,1} (\phi_3, \psi_0 [\phi_3])\right].
\label{vertex8}
\end{align}
We should note the following important fact:
By taking the on-shell limit, $\psi \longrightarrow \psi_0 [\phi_3] $, because the component $\Delta V^{1,0} (\phi_3 , \psi) /\Delta \psi $ in the first term goes to the differential coefficient $V^{1,1}(\phi_3, \psi_0[\phi_3])$, the terms (\ref{vertex8}) cancel each other out. 
So that, now, we have verified that the diagrams, less than 8 points, including some internal lines of ghost, do not contribute to the physical amplitudes.

Moreover, because the second term in (\ref{vertex8}) is the on-shell limit of the first term,
these terms could be expressed by using $\Delta$ defined in (\ref{Delta}),
i.e., 
\begin{align}
\left[ 
V^{1, 0}(\phi_3 , \psi) \frac{\Delta V^{1,0} (\phi_3 , \psi) }{\Delta \psi } 
-   V^{1,0}(\phi_3, \psi_0[\phi_3]) 
V^{1,1} (\phi_3, \psi_0 [\phi_3])\right] = \Delta \left[ V^{1,0} (\phi_3 ,\psi) 
\frac{\Delta V^{1,0} (\phi_3 , \psi )}{\Delta \psi }
{} \right] .
\end{align}
{We should note that the operation of the overall $\Delta$ also acts on $\Delta \psi$ .}
On the other hand,  $\Delta$ in the numerator of $\frac{\Delta V^{1,0} (\phi_3 ,\psi )}{\Delta \psi}$ should only be operated to $V^{1,0}$. Hence, more correctly, $\frac{\Delta V^{1,0} (\phi_3 ,\psi )}{\Delta \psi}$ is the term divided by $\Delta \psi$ after operating the $\Delta$ to $V^{1,0}$, i.e., we should represent it as follows, 
\begin{align}
\text{rhs} = \Delta \left[ V^{1,0} (\phi_3 , \psi) \frac{1}{\Delta \psi} \Delta V^{1,0} (\phi_3 , \psi) \right].
\end{align}

Now, we find that the terms proportional to $k^2$ are also equal to zero under the on-shell condition, and find also that these terms are proportional to $\Delta \psi$.
So that, these terms could also be eliminated by new field redefinition satisfying the expression (\ref{gg3}).
We could easily predict the appearance of the terms proportional to $k^3$ which are also equal to zero under the on-shell condition after the field redefinition.
Therefore, we could predict {the} realization of this mechanism in any order of $k$.
In the next part, we prove the realization of the mechanism.

\subsection{General Proof}
In the previous part, in lower orders of $k$, we saw that the first order terms with respect to $\lambda$ could be eliminated by the field redefinitions conserving the S matrix elements invariant.
In this part, we prove the possibility of the eliminations of the first order terms with respect to $\lambda$ in any order of $k$.
Then, by mathematical induction, the proof is performed for some general terms of the actions, the field redefinitions, and the $F^{(n)}$, given by the analogy of the previous part.
The general terms given by the analogy in the previous part are expressed as 
\begin{align}
&S_n[\phi_n, \psi ,\lambda ]=S_0[\phi_n ,\psi] +\int d^Dx 
\left[ (-1)^{n+1} \frac{k^{n-1}}{m^{2n-4}} \lambda  \Delta
\left( V^{1,0} (\phi_n,\psi)\frac{1}{\Delta \psi} \Delta \right)^{n-2}  V^{1,0} (\phi_n, \psi) \right.
\notag \\
& \left.  \ \ \ \ \ \ \ \ \ \ \ \ \ \ \ \ \ \ \ \ \ \ \ \ \ \ \ \ \ \ \ \ \ \ \ \ \ \ \ \ \ \ \
-\lambda \sum_{m=n}^{\infty} k^m F^{(m)}[\phi_n]
\right]+ \mathcal{O} (\lambda^2)  , \ \ \  (n\ge 2  ) ,   \label{kk10} \\
&\phi_{n} = \phi_{n+1} + \frac{(-1)^{n+1} k^{n-1}}{m^{2n-2}} \lambda\frac{1}{\Delta \psi} \Delta
\left( V^{1,0} (\phi_{n+1}, \psi )\frac{1}{\Delta \psi} \Delta \right)^{n-2} V^{1,0}(\phi_{n+1},\psi)
, \ \ \ (n\ge 2) , \label{kk9} \\
&F^{(n)}[\phi] = \lim_{\psi \longrightarrow \psi_0} \frac{(-1)^n}{m^{2n-2}} 
\left( V^{1,0} (\phi,\psi)
\frac{1}{\Delta \psi } \Delta  \right)^{n-1} V^{1,0} (\phi,\psi) . \ \ \ (n\ge 1) \label{kk7}
\end{align} 
{The operator $\Delta$ is defined as acting on all the functions in right hand side of $\Delta$.} 
An example is given by
\begin{align}
\left( V^{1,0}(\phi,\psi) \frac{1}{\Delta \psi} \Delta \right)^2  V^{1,0} (\phi,\psi)
&=   V^{1,0} (\phi, \psi)\frac{1}{\Delta \psi} \Delta  V^{1,0} (\phi,\psi)\frac{1}{\Delta \psi} \Delta V^{1,0}(\phi,\psi) \notag \\
&=  V^{1,0} (\phi,\psi)\frac{1}{\Delta \psi} \Delta  V^{1,0}(\phi,\psi) \frac{1}{\Delta \psi} 
\left(V^{1,0} (\phi , \psi)  -V^{1,0} (\phi, \psi_0[\phi]) \right) \notag \\
&= V^{1,0}(\phi, \psi) \frac{1}{\Delta  \psi} \left[ V^{1,0} (\phi , \psi) \frac{1}{\Delta \psi} 
\left(V^{1,0} (\phi , \psi)  -V^{1,0} (\phi, \psi_0[\phi]) \right) \right. \notag \\
&  \ \ \  \ \ \ \ \  \ \  \ \ \ \  \ \ \ \ \ \ \ \ 
- V^{1,0} (\phi, \psi_0[\phi]) V^{1,1}(\phi,\psi_0[\phi])  \biggl].
\end{align}
We should also note that the second or higher order terms with respect to the operator $ (1/\Delta \psi) \Delta $ are NOT asymptotic to the one just replaced the operator $ (1/\Delta \psi) \Delta $ to normal  derivative $d/d\psi$ in the limit of {the} on-shell condition.
The $n$-th order derivative of the function $f(\psi)$ is asymptotic to following term, 
\begin{align}
\lim_{\psi \longrightarrow \psi_0} \left( \frac{1}{\Delta \psi} \Delta \right)^n f(\psi) 
= \frac{1}{n!} \left[ \frac{d^n f (\psi) }{d\psi^n} \right]_{\psi= \psi_0}.\label{aa1}
\end{align}
There is a difference by the factor $1/n!$. 
{Furthermore}, the Leibniz rule is normally given by 
\begin{align}
\frac{1}{\Delta \psi } \Delta \left(f(\psi) g(\psi) \right)
 = \left( \frac{1}{\Delta \psi} \Delta f(\psi) \right)
\left( \lim_{\psi \longrightarrow \psi_0} g(\psi) \right) +
\left( \lim_{\psi \longrightarrow \psi_0} f(\psi) \right)  \left( \frac{1}{\Delta \psi} \Delta g(\psi) \right).\label{aa2}
\end{align}
By using (\ref{aa1}) and (\ref{aa2}), we can check the equations (\ref{kk10})-(\ref{kk7}) straightforwardly.

Now, let us prove the equations (\ref{kk10})-(\ref{kk7}).
First of all, we should note that the recursion equations (\ref{kk9}) for $\phi_n$ are not the proposition that should be proved, but the given redefinitions which define the functional forms of the actions $S_n$. In other words, the functional $S_n[\phi_n{,} \psi, \lambda ]$ is defined by the recursions $S_{n-1}[\phi_{n-1} [\phi_n] , \psi ,\lambda ]=S_n[\phi_n, \psi, \lambda]$ with the initial term (\ref{jj2}). 
In order to verify the correctness of the actions (\ref{kk10}), it is necessary to verify the correctness of the equations (\ref{kk7}) in advance.

Let us verify the correctness of the equations (\ref{kk7}).
For this purpose, we again solve the equation (\ref{kk2}) which $F[\phi]\equiv \sum_{n=1}^{\infty} k^n F^{(n)}[\phi]$ should {satisfy}, by using the operator $\Delta$.
The form of the equation (\ref{kk2}) is now given by
\begin{align}
F[\phi] + k V^{1,0}\left( \phi, \psi_0[\phi] + \frac{F[\phi]}{m^2}  \right) =0. \label{kk6}
\end{align} 
By decomposing the term $V^{1,0} \left(\phi, \psi_0[\phi] + \frac{1}{m^2} F \right)$ in the terms which do not 
include any $k$ and the other terms, we obtain
\begin{align}
&F[\phi] = -k \left[ V^{1,0}(\phi, \psi_0[\phi] )
+ \Delta V^{1,0} \left(\phi,  \psi_0[\phi] + \frac{F[\phi]}{m^2}  \right) \right], \notag \\
&\Delta V^{1,0} \left(\phi,  \psi_0[\phi] + \frac{F[\phi]}{m^2}  \right) \equiv 
 V^{1,0} \left(\phi,  \psi_0[\phi] + \frac{F[\phi]}{m^2} \right) 
- \lim_{F\longrightarrow 0}  V^{1,0} \left( \phi,  \psi_0[\phi] + \frac{F[\phi]}{m^2}  \right) .\label{aa4}
\end{align}
Here, the definition of operator $\Delta$ is identical with the one used in (\ref{Delta}) or (\ref{Delta2}) 
in the previous section. 
Now, the first term $V^{1,0}(\phi,\psi_0[\phi])$ {becomes} independent of $k$,
and the second term $\Delta V^{1,0}$ {includes} the first or higher order terms
with respect to $k$.
Then, the first order term $F^{(1)}[\phi]$ has been decided.
In this way, we can pick up the lowest order terms with respect to $k$ by using the operator $\Delta$.

Moreover, the terms $\Delta V^{1,0}$ could be decomposed as follows, 
\begin{align}
\Delta V^{1,0} \left( \phi,  \psi_0 + \frac{F}{m^2}  \right) 
&=  \frac{F}{m^2} \frac{1}{F/m^2}
\Delta V^{1,0} \left( \phi,  \psi_0 + \frac{F}{m^2}  \right) \notag \\
&=-\frac{k}{m^2}  V^{1,0} \left(\phi,  \psi_0 + \frac{F}{m^2}  \right)
 \frac{1}{F/m^2}
\Delta V^{1,0} \left( \phi,  \psi_0 + \frac{F}{m^2}  \right) \notag \\
&= -\frac{k}{m^2} \left[ \lim_{F\longrightarrow 0}
 V^{1,0} \left(\phi,  \psi_0 + \frac{F}{m^2}  \right)
 \frac{1}{F/m^2}
\Delta V^{1,0} \left( \phi,  \psi_0 + \frac{F}{m^2}  \right) \right. \notag \\
& \left. \ \ \ \ \ \ \ +
\Delta V^{1,0} \left(\phi,  \psi_0 + \frac{F}{m^2}  \right)
 \frac{1}{F/m^2}
\Delta V^{1,0} \left( \phi,  \psi_0 + \frac{F}{m^2}  \right)
  \right].
\end{align}
Here, just for simplicity, we omit the dependence $[\phi]$ in the expressions of $\psi_0$ and $F$.
In the second line, we substitute the equation (\ref{kk6}) {into} the component $F$ in the numerator. 
In the third line, the first term {becomes} the lowest order term with respect to $k$,
and the second term {includes} the higher terms.
Then the second order term $F^{(2)}$ {has} been decided.

In this way, the term $F^{(n)}[\phi]$ could be decided in order by order, by using the equation (\ref{kk6}) and $\Delta$. 
In the same way as the above procedure, we obtain general recursion equations,
\begin{align}
&\Delta \left\{ V^{1,0} \left( \phi , \psi_0 + \frac{F}{m^2} \right)\frac{1}{F/m^2} \Delta \right\}^n 
 V^{1,0} \left( \phi , \psi_0 + \frac{F}{m^2} \right)  \notag \\
&= -\frac{k}{m^2}  V^{1,0} \left(\phi,  \psi_0 + \frac{F}{m^2} \right) \frac{1}{F/m^2}
\Delta \left\{ V^{1,0} \left( \phi , \psi_0 + \frac{F}{m^2} \right) \frac{1}{F/m^2} 
\Delta \right\}^n 
 V^{1,0} \left( \phi , \psi_0 + \frac{F}{m^2} \right)  \notag \\
&=  -\frac{k}{m^2} \left[ \lim_{F\longrightarrow 0} 
 \left\{ V^{1,0} \left( \phi , \psi_0 + \frac{F}{m^2} \right) \frac{1}{F/m^2} 
\Delta \right\}^{n+1} 
 V^{1,0} \left( \phi , \psi_0 + \frac{F}{m^2} \right) \right. \notag \\
& \left. \ \ \ \ \ \ \ \ \ \ \
 +\Delta  \left\{ V^{1,0} \left( \phi , \psi_0 + \frac{F}{m^2} \right)\frac{1}{F/m^2} \Delta \right\}^{n+1} 
 V^{1,0} \left( \phi , \psi_0 + \frac{F}{m^2} \right)
\right] {.}
\end{align}
By substituting this recursion equations {into} the equation (\ref{aa4}) in order by order, 
we obviously obtain the complete form of $F$ as follows, 
\begin{align}
F&= \sum_{n=1}^{\infty} \lim_{F \longrightarrow 0 }  \frac{(-k)^n}{m^{2n-2}} \left( V^{1,0}
 \left( \phi,\psi_0+\frac{F}{m^2} \right)
\frac{1}{ F/m^2 } \Delta  \right)^{n-1} V^{1,0} \left(\phi, \psi_0 + \frac{F}{m^2} \right) \notag \\
&= \sum_{n=1}^{\infty} \lim_{\psi \longrightarrow \psi_0 }  \frac{(-k)^n}{m^{2n-2}} \left( V^{1,0}
 \left( \phi,\psi \right)
\frac{1}{\Delta \psi } \Delta  \right)^{n-1} V^{1,0} \left(\phi, \psi \right) .
\end{align}
Therefore, we have verified the equation (\ref{kk7}).

Finally, let us complete our proof by confirming the correctness of general terms of the action (\ref{kk10}) by the mathematical induction. 
In the case of $n=2$, it is obvious that {the action (\ref{kk10}) coincide with} (\ref{jj2}).
Now, for the fixed $n$, we assume that the equation (\ref{kk10}) is correct. 
Under the the field redefinition (\ref{kk9}), it is obvious from the equation (\ref{kk8}) that
the original action $S_0[\phi_n , \psi]$ of the equation (\ref{kk10}) transforms as follows, 
\begin{align}
S_0[\phi_n, \psi] - S_0[\phi_{n+1} , \psi] &= - \int d^Dx [m^2 \Delta \psi + k V^{1,0}(\phi_{n+1},\psi)] \notag \\
& \ \ \ \ \ \ \times \frac{(-1)^{n+1} k^{n-1}}{m^{2n-2}} \lambda\frac{1}{\Delta \psi} \Delta
\left( V^{1,0} (\phi_{n+1},\psi) \frac{1}{\Delta \psi} \Delta \right)^{n-2} V^{1,0}(\phi_{n+1},\psi) 
 + \mathcal{O}(\lambda^2) \notag \\
&=\int d^Dx\left[ - \frac{(-1)^{n+1} k^{n-1}}{m^{2n-4} } \lambda \Delta
\left( V^{1,0} (\phi_{n+1},\psi) \frac{1}{\Delta \psi} \Delta \right)^{n-2} V^{1,0}(\phi_{n+1},\psi) 
\right. \notag \\
& \left. \ \ \ \ \ \ + \frac{(-1)^{n+2} k^{n}}{m^{2n-2}}  \lambda
\left( V^{1,0}(\phi_{n+1},\psi)  \frac{1}{\Delta \psi} \Delta \right)^{n-1} V^{1,0} (\phi_{n+1},\psi) \right]  + \mathcal{O}(\lambda^2).
\end{align}
Then the $(n-1)$-th order terms with respect to $k$ have been eliminated as follows, 
\begin{align}
S_{n+1}[\phi_{n+1}, \psi] \equiv S_n[\phi_n[\phi_{n+1}] ,\psi] =& 
 S_0[\phi_{n+1} ,\psi ] \notag \\
&+  \int d^Dx \left[\frac{(-1)^{n+2} k^{n}}{m^{2n-2}}  \lambda
\left( V^{1,0} (\phi_{n+1},\psi) \frac{1}{\Delta \psi} \Delta \right)^{n-1} V^{1,0} (\phi_{n+1},\psi)
 \right. \notag \\
& \left. -\lambda k^n F^{(n)}[\phi_{n+1}] -\lambda \sum_{m=n+1}^{\infty} k^m F^{(m)}[\phi_{n+1}]\right] + \mathcal{O}(\lambda^2).
\end{align}
By using the equation (\ref{kk7}) which has already been proved, the $n$-th order terms with respect to $k$ are expressed as follows, 
\begin{align}
&\frac{(-1)^{n+2} k^{n}}{m^{2n-2}}  \lambda\left[
\left( V^{1,0} (\phi_{n+1},\psi) \frac{1}{\Delta \psi} \Delta \right)^{n-1} V^{1,0} (\phi_{n+1},\psi) -
\lim_{\psi \longrightarrow \psi_0} \left( V^{1,0} (\phi_{n+1},\psi) \frac{1}{\Delta \psi} \Delta \right)^{n-1} V^{1,0}  (\phi_{n+1},\psi) \right] \notag \\
&=\frac{(-1)^{n+2} k^{n}}{m^{2n-2}}  \lambda \Delta \left( V^{1,0} (\phi_{n+1},\psi) \frac{1}{\Delta \psi} \Delta \right)^{n-1} V^{1,0} (\phi_{n+1},\psi).
\end{align}
Finally, we obtain the $(n+1)$-th action, 
\begin{align}
S_{n+1}[\phi_{n+1},\psi] =& S_0[\phi_{n+1},\psi] \notag \\
&+ \int d^D x \left[
 \frac{(-1)^{n+2} k^{n}}{m^{2n-2}}  \lambda \Delta \left( V^{1,0}  (\phi_{n+1},\psi) \frac{1}{\Delta \psi} \Delta \right)^{n-1} V^{1,0}  (\phi_{n+1},\psi) \right. \notag \\
& \left. \ \ \ \ \ \ \ \ \ \  \ \ \ \ \ \ \ \ \ \ \ \ \ \ \ \ \ \ \ \ \ \ \
 -\lambda \sum_{m=n+1}^{\infty} k^m F^{(m)}[\phi_{n+1}] 
\right]+ \mathcal{O}(\lambda^2).
\end{align}
Therefore, we have completed the proof by using the mathematical induction.

Because the above proof has been a little bit complicated, we now summarize how the conjecture have been proved. By using Eq.~(\ref{kk10}), in the limit $n\longrightarrow \infty$,
we obtain the following expression,
\begin{align}
S_\infty [\phi_\infty, \psi ,\lambda ]=S_0[\phi_\infty ,\psi] + \mathcal{O}(\lambda^2). \label{nn1}
\end{align}
From the argument of section \ref{asc}, the terms $\mathcal{O}(\lambda^2)$ could not contribute to any physical amplitude. The interaction terms which contribute to the physical amplitudes 
are only given by the original interaction term $V(\phi, \psi)$.
Therefore, in this frame, it is obvious that the physical amplitudes of the higher derivative theory coincide with the corresponding amplitudes of the original theory.

Moreover, this frame {is related} with the initial frame obtained by (\ref{jj2}) through the field redefinitions (\ref{kk9}). Because the field redefinitions (\ref{kk9}) satisfy the expression (\ref{gg3}), any scattering amplitudes calculated in the initial frame (\ref{jj2}) {coincide} with the amplitudes of the asymptotic frame (\ref{nn1}). 

So that, through the asymptotic frame, the physical amplitudes calculated in the initial frame coincide with the corresponding amplitudes in original theory. Because the conjecture (\ref{ex1}) is given in the initial frame (\ref{jj2}), [that is, (\ref{sc4}) in Section \ref{ms}], the conjecture has been proved.

\section{SUMMARY}
In this paper, we have investigated the possibility of the elimination of the ghost in the higher derivative theory proposed in \cite{Hassan1}.
Although the possibility of the elimination of the ghost in the linear level had been argued by Hassan {\it et al.} \cite{Hassan1}, it is not so trivial to check whether the ghost could be eliminated in the non-linear level or not.
We have considered the model with two scalar fields interacting with each other by a mass mixing (\ref{sc1}), which was proposed in \cite{Hassan1}, but in the analysis of  \cite{Hassan1}, the {non-linear} interaction terms were neglected.
We have analyzed the model without neglecting the interaction terms.
In Section \ref{111}, although there are many algebraic solutions, we have adopted the unique 
solution $\psi[\phi]$ which satisfy the condition $\psi[\phi =0] =0$.
Under this assumption, we have found that in the higher derivative theory, there appears a ghost mode in addition to two healthy modes corresponding to the modes in the original theory.
We have called these healthy modes as ``physical fields''.
In Section \ref{ex}, we have defined ``physical amplitudes'' as the amplitudes where all the external lines are taken to ``physical fields''.
We have also proposed the conjecture (\ref{ex1}), where ``physical amplitudes'' of the higher derivative theory coincide with the amplitudes of the original theory.
In this setup, we have proved the conjecture (\ref{ex1}) without any additional assumption besides the ones given  in Section \ref{333}.

It could be straightforward to extend the analyses given in this paper to the bigravity theory. 
There is, however, one concern. 
In order to apply the {arguments in this} paper to the bigravity theory, we need to pay more attentions to the commutativity between the substitution of the algebraic solution and the procedure of the gauge fixing.
Because, as argued in the Appendix \ref{aa}, the procedure of {the} derivation of the higher derivative theory could be regarded as the equivalent rewriting of the path integral.
In the path integral formulation, we integrate out the field $\phi$ first. In the case of the gauge theory, we cannot perform the integral without the gauge fixing. Then we should fix the gauge first of all. 
Therefore, it could be necessary to investigate the commutativity between the substitution of the algebraic solution and the procedure of the gauge fixing.

In order to investigate this problem if the commutativity, it could be also better to investigate some toy model first. 
There is a candidate of the toy model: 
The pseudo-linear model \cite{Hinterbichler:2013eza} is the massive spin-2 model which has the non-derivative interaction terms in addition to the linear terms in the Fierz-Pauli model.
Moreover, by Hinterbicher in \cite{Hinterbichler:2013eza}, it has been proved that the BD ghost does not 
appear in this model. 
The curved space extension of the proof has been given in \cite{new curved}-\cite{Ohara:2014vua}.
By using the pseudo-linear model and the linearized Einstein-Hilbert action, we easily construct the model with mass mixing such as the bigravity.
This theory has the structure very {similar} to that in the $\xi^3$-model which have used in section \ref{asc}.
By this future work, we may investigate the commutativity between the substitution of the algebraic solution and the procedure of the gauge fixing.

\appendix

\section{PATH INTEGRAL}\label{aa}
In order to show the correspondence between the higher derivative theory and the original theory, 
we consider how the higher derivative theory is obtained by the equivalent transformations of the generating  function of the original theory.
Let us consider the path integral with the external sources $J_\phi, J_\psi$, 
\begin{align}
Z[J_\phi, J_\psi] = \int D \phi D\psi \exp \left[i S_0[\phi,\psi] + i\int d^D x \left( \phi J_\phi +\psi J_\psi \right) \right]. \label{app1}
\end{align}
By integrating out $\phi$, we obtain 
\begin{align}
Z[J_\phi, J_\psi] = \int D\psi \exp \left[i S_0[\phi[\psi, J_\phi ],\psi] + i\int d^D x \left( \phi [\psi, J_\phi] J_\phi+\psi J_\psi \right) \right].
\end{align}
Here, in the tree level, {$\phi[\psi, J_\phi ]$} is defined as the perturbative solution of the equations of motion,
\begin{align}
\frac{\delta S_0 [\phi, \psi]}{\delta \phi} + J_{\phi} =0, \label{mm1}
\end{align}
with respect to $\phi$ around the vacuum $\phi= 0=\psi$.
Around this vacuum, the unique inverse function $\psi[\phi, J_\phi]$, which is the algebraic solution of the equation (\ref{mm1}) with respect to $\psi$ with the condition $\psi[\phi=0, J_{\phi}=0] =0 $ , exists. Under the field redefinition $\psi = \psi[\phi, J_\phi]$,
because the inverse function $\psi[\phi, J_\phi]$ satisfies the identity $\phi[\psi[\phi , J_\phi ], J_\phi]= \phi$, we obtain
\begin{align}
Z [J_\phi, J_\psi]= \int D \phi \text{Det}\left[ 
\frac{\delta \psi[\phi, J_\phi ]}{\delta \phi} \right] \exp \left[i S_0[\phi,\psi[\phi, J_\phi ]] + i\int d^D x \left( \phi J_\phi+\psi [\phi, J_\phi] J_\psi\right) \right] . \label{app2}
\end{align}
By introducing the Lagrange multiplier field $\lambda$ and the FP-ghosts $C$, $\bar{C}$,
we obtain
\begin{align}
Z[J_\phi, J_\psi] = \int D \phi D \psi D \lambda DC D\bar{C} \exp \left[ i S_0[\phi,\psi] + i\int d^D x \left(
m^2 \lambda \left( \psi -\psi[\phi, J_\phi] \right)+ i \bar{C} \frac{\delta \psi[\phi, J_\phi ]}{ \delta \phi} C+ \phi J_\phi +\psi J_\psi \right) \right]. \label{mm2}
\end{align}
Here, we omit an integral for the FP-ghosts.
The exponent of this expression is similar to that in (\ref{gg5}) with source terms for $\phi$ and $\psi$.
The different parts are the terms $\psi[\phi , J_\phi]$ in the Lagrange multiplier terms and the FP-ghosts terms.

Now, we consider the specific case where {two fields interacted with each other only through the mass mixing term}, i.e.,{ 
$V(\phi, \psi)$=$V_\phi(\phi)+V_\psi(\psi)$}.
In this case, the solution of the equation (\ref{mm1}) is expressed as follows,
\begin{align}
m^2 \psi[\phi , J_\phi ] = m^2 \psi[\phi] + J_\phi, \ \ \ \ \psi[\phi] \equiv \psi[\phi, 0].
\end{align}
By substituting this expression into the path integral (\ref{mm2}), we obtain
\begin{align}
Z[J_\phi, J_\psi] = \int D \phi D \psi D \lambda DC D\bar{C} \exp \left[ i S_0[\phi,\psi] + i\int d^D x \left(
m^2 \lambda \left( \psi -\psi[\phi] \right)+ i \bar{C} \frac{\delta \psi[\phi]}{ \delta \phi} C+ (\phi- \lambda) J_\phi +\psi J_\psi \right) \right]. \label{mm2BB}
\end{align}
In the tree level, we could ignore the FP-ghost terms. Under the field redefinition $\phi \longrightarrow \phi + \lambda $, we obtain 
\begin{align}
Z[J_\phi, J_\psi] & {\approx} \int D \phi D \psi D \lambda \exp \left[ i S_0[\phi + \lambda ,\psi] + i\int d^D x \left(
m^2 \lambda \left( \psi -\psi[\phi+\lambda] \right)+ \phi J_\phi +\psi J_\psi \right) \right]
\notag \\
&=\int D \phi D \psi D \lambda \exp \left[ i S[\phi + \lambda ,\psi, \lambda] + i\int d^D x \left( \phi J_\phi +\psi J_\psi \right) \right]
. \label{mm2CC}
\end{align}
Here, {the equal ``$\approx$" means the equivalence up to the FP-ghost terms,} the action $S[\phi+\lambda, \psi, \lambda]$ is defined in (\ref{gg5}), and the linear part of 
$S[\phi+\lambda, \psi, \lambda]$ is given in (\ref{gg1}).
By the fields redefinition (\ref{mm3}), we obtain the correspondence between the Green functions,
\begin{align}
\int D \xi D\eta \exp \left[i S_0[\phi(\xi, \eta),\psi (\xi, \eta )] + i\int d^D x \left( \xi J_\xi +\eta J_\eta \right) \right]
&{\approx} \int D \xi D \eta D \lambda \exp \left[ i \bar{S}[\xi, \eta, \lambda] + i\int d^D x \left( \xi J_\xi +\eta J_\eta \right) \right], \notag \\
J_\xi \equiv \frac{1}{\sqrt{2}} \left( J_\phi + J_\psi \right)&, \ \ \ \ \
J_\eta \equiv \frac{1}{\sqrt{2}} \left( J_\phi - J_\psi \right).
\end{align}
Here, the $\bar{S}[\xi, \eta, \lambda]$ is given in the equation (\ref{sc4}). 
Therefore, in the case of {$V(\phi, \psi) =V_\phi(\phi) + V_\psi(\psi)$}, the conjecture (\ref{ex1}) 
can be trivially shown.

In the case of {$V(\phi, \psi) \neq V_\phi(\phi) + V_\psi(\psi)$}, however, the correspondence is not so trivial, due to the non-linear dependence of $J_\phi$ in the equation (\ref{mm2}). 
These terms contribute to the Green functions as some composite fields.
In order to confirm the conjecture (\ref{ex1}), we should show that the diagrams with such composite fields vanish under the on-shell condition.
We do not, however, continue the further analysis by using the path integral but we prove the conjecture in another way in this paper. 

Now, let us compare the above argument with the argument by Hassan {\it et al.} in \cite{Hassan1}.
They also considered the model with the source terms (\ref{app1}), but they did not argue by using the integration as given above. Their arguments were more straightforward. 
First, they straightforwardly calculated the algebraic solution of Eq.~(\ref{mm1}) in the case of $V(\phi, \psi) =0 $, and substituted the solution to the original action (\ref{app1}), which coincides with Eq.~(\ref{app2}) without using the Jacobian.
After that, by integrating out the obtained higher derivative theory, they obtained the generating function identical with the original theory.
They also commented on the extension to the case of {$V(\phi, \psi)$=$V_\phi(\phi)+V_\psi(\psi)$}. 
They claimed the equivalence of both theories based on {the} above arguments, so they have not shown the correspondence in the case of {$V(\phi, \psi) \neq V_\phi(\phi) + V_\psi(\psi)$}. This is our motivation for considering the non-linear case.

\end{document}